\documentclass[prl,twocolumn,superscriptaddress]{revtex4-2}

\usepackage{graphicx}  
\usepackage{amsmath}   
\usepackage{amssymb}   
\usepackage[colorlinks,linkcolor=blue,anchorcolor=pink,filecolor=green,menucolor=yellow,citecolor=blue,urlcolor=blue,bookmarks=false,hypertexnames=true]{hyperref} 
\usepackage{natbib}    

\usepackage[normalem]{ulem}
\usepackage{lineno}
\usepackage[usenames,dvipsnames]{xcolor}

\begin{document}

\title{Collective ballistic motion explains fast aggregation in adhesive active matter}
\author{Emanuel F.~Teixeira}
\email{emanuel.teixeira@ufrgs.br}
\affiliation{Instituto de Física, Universidade Federal do Rio Grande do Sul, CP 15051, CEP 91501-970 Porto Alegre - RS, Brazil}%

\author{P.~de Castro}
\email{pablo.castro@unesp.br}
\affiliation{ICTP South American Institute for Fundamental Research, Instituto de Física Teórica, UNESP -- Universidade Estadual Paulista,
Rua Dr.~Bento T.~Ferraz 271, 01140-070, São Paulo, SP, Brazil}%

\author{Carine P.~Beatrici}
\email{carine@if.ufrgs.br}
\affiliation{Instituto de Física, Universidade Federal do Rio Grande do Sul, CP 15051, CEP 91501-970 Porto Alegre - RS, Brazil}%

\author{Heitor C.~M.~Fernandes}
\email{heitor.fernandes@ufrgs.br}
\affiliation{Instituto de Física, Universidade Federal do Rio Grande do Sul, CP 15051, CEP 91501-970 Porto Alegre - RS, Brazil}%

\author{Leonardo G.~Brunnet}
\email{leon@if.ufrgs.br}
\affiliation{Instituto de Física, Universidade Federal do Rio Grande do Sul, CP 15051, CEP 91501-970 Porto Alegre - RS, Brazil}%

\date{\today}

\begin{abstract}
Inspired by motile cells in tissue formation, we find that active systems of self-aligning adhesive particles undergo ballistic aggregation through a flocking transition. This kinetic regime emerges when the cluster persistence length grows faster with cluster mass than the intercluster distance does.  We also identify and explain distinct non-collective kinetic regimes, including biologically relevant long-lived transients. Our analytical and numerical results offer a unified framework explaining the broad range of experimentally observed aggregation exponents in cellular systems and reveal physical principles potentially critical for timely tissue organization.
\end{abstract}

\maketitle

{\it Introduction}---Aggregation processes are ubiquitous, from aerosols to living matter~\cite{xiong2001morphological,hyman2014liquid,franklin2016handbook,de2023sequential,caporusso2023dynamics}. In active systems, motile cells can agglomerate, driving phenomena like bacterial biofilm initiation~\cite{grobas2021swarming,o2000biofilm} and tissue formation~\cite{heer2017tension}. In particular, there has been long-standing interest in cell-sorting experiments~\cite{krens2011cell,teixeira2025segregation,bothe2025eph}, where cells clusters meet and grow by successive mergers~\cite{krens2011cell,teixeira2025segregation,bothe2025eph}. In these systems, strong cell–cell adhesion suppresses cluster fragmentation. Yet, complex interactions and motion generate rich aggregation kinetics that remain elusive despite their key role in timely tissue organization~\cite{mehes2012collective}.

In diffusion-limited cluster aggregation (DLCA), clusters move diffusively with a diffusivity that decreases inversely with cluster mass \cite{leyvraz2003scaling}. This classical picture predicts that in two dimensions, without cluster fragmentation, the average cluster mass grows as \( \overline{M}(t) \sim t^z \) with an aggregation exponent \( z = 1/2 \)~\cite{nakajima2011kinetics,cremer2014scaling,durand2021}. In contrast, ballistic cluster motion yields \( z \simeq 1\), provided that self-propulsions are uncorrelated between cells~\cite{jiang1993scaling,leyvraz2003scaling,krapivsky2010kinetic}. However, experiments in active tissues report faster coarsening dynamics with \( z > 1 \)~\cite{mehes2012collective}, indicating that classical models are insufficient.

Motivated by these discrepancies, Mones et al.~\cite{mones2015anomalous} simulated adhesive active particles with self-alignment \cite{baconnier2025self}, where self-propulsion gradually aligns with the particle's net velocity, obtaining $z \simeq 2$. This rapid coarsening was attributed to collective motion, akin to Vicsek-like systems~\cite{belmonte2008,paul2024finite,cremer2014scaling}, but its mechanistic origin remained unexplained. Subsequently, Beatrici et al.~\cite{beatrici2017mean} showed how internal cluster alignment alters the diffusivity-mass relation, modifying $z$; however, their analysis was limited to weak alignment and persistence, yielding at most $z=1$.

In this work, we show that the anomalous exponent $z \simeq 2$ arises because, as a consequence of flocking, the cluster persistence length grows linearly with cluster mass, while the intercluster distance scales as the square root of mass. This corresponds to a crossover into the regime termed here collective ballistic aggregation (CBA).
Our analysis relies on simulations of adhesive active particles, in which we varied both particle persistence and self-alignment strength. Beyond CBA, we also elucidate other mechanisms underlying a broad range of experimental aggregation exponents. Our results are supported by analyses of single-cluster dynamics and a generalized Smoluchowski coagulation theory \cite{smoluchowski1916drei}. The present work unifies disparate experimental and theoretical observations and shows how cooperative behavior reshapes aggregation kinetics in living matter.
 \begin{figure*}[!ht]
    \centering
\raisebox{3mm}{\includegraphics[width=0.31\textwidth]{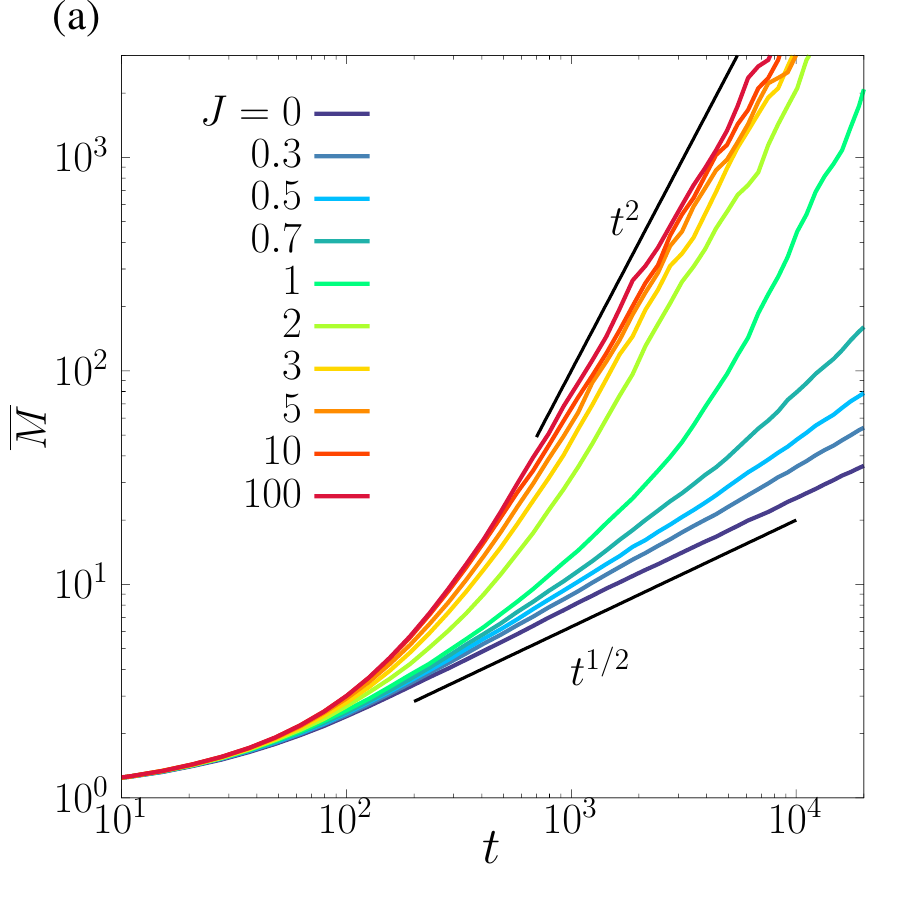}}
 \hspace{0.3cm}
\includegraphics[width=0.62\textwidth]{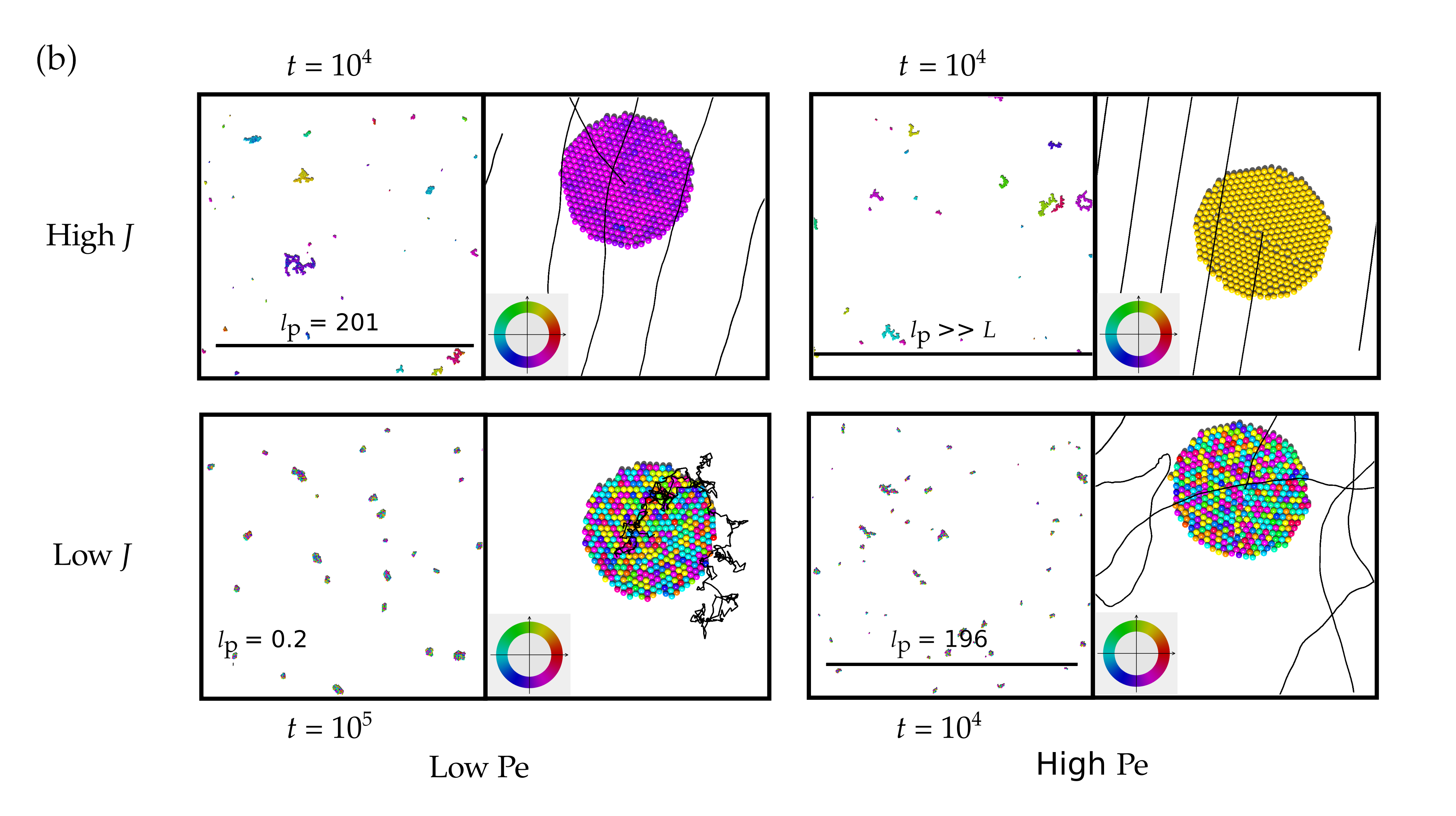}
 \vspace{0.0cm} 
\caption{Adhesive active matter exhibits rich aggregation kinetics. (a) Time evolution of the average cluster mass $\overline{M}(t)$ for different values of $J$. Stronger alignment accelerates aggregation and increases exponents. Here, $\mathrm{Pe} = 1$. (b) Snapshots of aggregation and cluster dynamics in different regimes for low alignment ($J = 0$), high alignment ($J = 100$), low individual-particle persistence ($\mathrm{Pe} = 1$) and high individual-particle ($\mathrm{Pe} = 10^{3}$). Scale bars show cluster persistence length. Colors indicate particle polarity (see color wheel). $L$ is the simulation box length.
}
 \label{fig1}
\end{figure*}

{\it Model}---We employ a minimal model inspired by motile tissue cells~\cite{szabo2006phase} with particles represented as disks subject to contact forces and self-alignment: each particle self-propels along a noisy direction that gradually aligns with its own net velocity. A strong attractive force accounts for cell adhesion. Self-alignment induces collective motion and is more realistic in this context than the original Vicsek alignment \cite{vicsek1995novel}, since tissue cells cannot explicitly average over neighbors \cite{szabo2006phase}. Our results are robust to the choice of alignment rule, as observed for Vicsek-like alignment at low strength \cite{beatrici2017mean}.

We consider $N$ adhesive active Brownian disks in 2D with periodic boundary conditions~\cite{szabo2006phase,henkes2011active}. Their dynamics is given by
\begin{eqnarray}
\dot{\mathbf{r}}_i &=& v_{0}\, \mathbf{n}_{i} + \mu\sum_{j \neq i} \mathbf{F}_{i j} \,, \\
\dot{\theta}_i &=&  - J\frac{\partial }{\partial \theta_i} \left ( \mathbf{n}_{i}\cdot\hat{\mathbf{v}}_{i} \right ) + \sqrt{2 D_{R}} \, \xi_{i}\,,
\end{eqnarray}
where $v_0$ is the self-propulsion speed, $\mathbf{n}_i = (\cos \theta_i, \sin \theta_i)$ is the self-propulsion direction, also called polarity, $\xi_i$ is a Gaussian white noise with $\langle \xi_i(t)\xi_j(t') \rangle = \delta_{ij}\delta(t-t')$ and $\hat{\mathbf{v}}_i = \mathbf{v}_i / |\mathbf{v}_i|$, where $\mathbf{v}_i=\dot{\mathbf{r}}_i$. 
Parameters $D_R$ and $J$ set the reorientation timescales $\tau_R = 1/D_R$ and $\tau = 1/J$, respectively. 
The self-alignment term has a pseudo potential $E_{\rm align} = -J\, \mathbf{n}_i\cdot\hat{\mathbf{v}}_i$~\cite{sarkar2021minimal,barton2017active}. 
The force $\mathbf{F}_{ij}$ between particles $i$ and $j$ comprises a short-range repulsion and a short-range adhesion. 
For interparticle distances $|\mathbf{r}_{ij}| \leq \sigma$, a repulsive force $\mathbf{F}_{ij}^{\rm rep} = -k_{c} \left(|\mathbf{r}_{ij}| - \sigma\right)\hat{\mathbf{r}}_{ij}$ accounts for overlap between the disks, where $\sigma$ is their diameter and $\hat{\mathbf{r}}_{ij} = \mathbf{r}_{ij} / |\mathbf{r}_{ij}|$ is the unit vector connecting the centers of particles $i$ and $j$. 
In the range $\sigma < |\mathbf{r}_{ij}| \leq l_{\rm adh}$, an adhesive force $\mathbf{F}_{ij}^{\rm adh} = -k_{\rm adh} \left(|\mathbf{r}_{ij}| - \sigma\right)\hat{\mathbf{r}}_{ij}$ attracts neighboring particles, where $l_{\rm adh}$ sets the adhesive interaction range.



We measure length and time in units of $\sigma$ and $\tau_{0} = \sigma/v_{0}$, kept constant throughout this work. 
We thus express the polarity decorrelation time  $\tau_{R}$ as the (rotational) Péclet number~\cite{martin2018collective,zhao2021phases}, $\mathrm{Pe} \equiv v_0 \tau_{R}/\sigma = \tau_{R}/\tau_{0}$. We start from a random initial condition and let the system evolve into growing clusters by merger events where particles irreversibly adhere. Unless otherwise specified, we employ $N=4\times10^{4}$ particles. To suppress multi-cluster mergers and percolation effects, we use a low packing fraction, $\phi = 0.01$. This produces slow aggregation kinetics (without altering the exponents) in which the cluster-mass distribution evolves more smoothly and the power-law regime is extended in time, thereby improving statistics. We also performed tests at moderately higher $\phi$ and observed the same exponents. Additional details are in the Supplemental Material \cite{supp}. 

{\it Low particle persistence}---First, we consider self-alignment in the low-$\mathrm{Pe}$ regime ($\mathrm{Pe} = 1$). 
To quantify the aggregation kinetics, Figure~\ref{fig1}a shows our results for the average cluster ``mass'' (i.e., the number of particles in a cluster) versus time, where a power-law behavior, $\overline{M}(t) \sim t^{z}$, sets in after some transient time. 
Between $J=0$ and $J \approx 0.7$, the aggregation exponent changes from $z=1/2$ to $z=1$. 
Upon further increasing $J$, $z$ jumps to $z \approx 2$. Snapshots of the aggregates are shown in Figure~\ref{fig1}b; for movies, see Supplemental Material \cite{supp}. Later in this work, a comparison between intercluster distance and cluster motion persistence length will be motivated and presented.

\begin{figure}[!h]
\centering
\includegraphics[width=1.0\columnwidth]{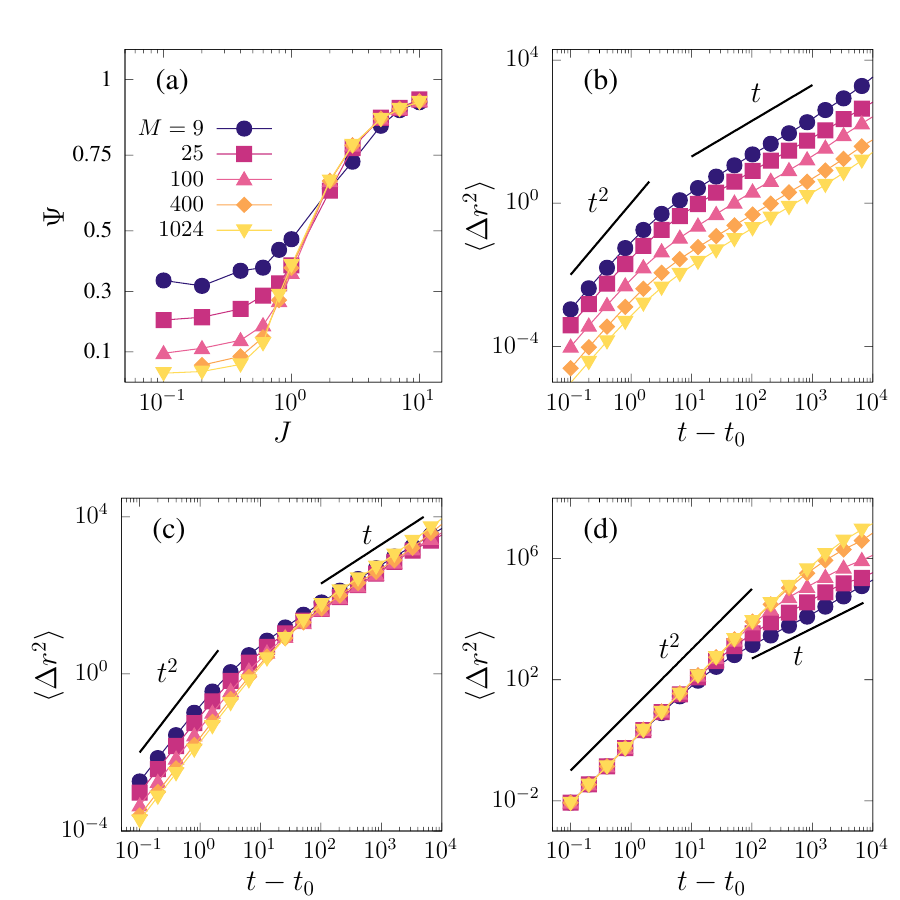}
\caption{Self-alignment induces a collective motion transition, altering the cluster ballistic regime duration and speed as well as the effect of cluster mass on motion.
(a) Collectivity order parameter $\Psi$ versus $J$ for different masses. A transition  occurs near $J \approx 0.6$. MSD curves, $\left< \Delta r^{2}\right>(t)$, for (b) the disordered regime ($J = 0$), (c) near the transition ($J = 0.6$), and (d) in the collective regime ($J = 10$). Here,  $\mathrm{Pe} = 1$.}
\label{fig2}
\end{figure}

To understand the dependence of $z$ on $J$, we simulated single isolated clusters since their movement affect $z$. Cluster trajectories are shown also in Figure~\ref{fig1}b; for movies of single-cluster simulations, see Supplemental Material \cite{supp}.
We quantify the alignment of polarities within a cluster using the global polar order parameter, 
${\Psi = \left< \left | \sum_{i}^{M} \mathbf{n}_{i}\right |\right>_{t}/M}$, where $\left< ...\right>_{t}$ 
is the steady-state average \cite{martin2018collective}. High ($\Psi \to 1$) or low ($\Psi \to 0$) collectivity reflects strong or weak alignment, respectively.
Figure \ref{fig2}a displays $\Psi$ versus $J$ for different cluster masses, showing that $J$ induces polar order \cite{vicsek1995novel,szabo2006phase,peruani2012collective,martin2018collective}.  Polar order emerges as particles adhere and move together with similar velocities. At strong $J$, particles eventually self-align with their common velocity. The difference in $\Psi$ between order and disorder increases with mass.

To characterize single-cluster motion, we compute the mean-squared displacement (MSD), $\left< \Delta r^{2}\right> = \left< \left ( \mathbf{r}_\mathrm{cm}(t) - \mathbf{r}_\mathrm{cm}(t_{0})\right )^{2}\right>$, where $\mathbf{r}_\mathrm{cm}$ is the cluster center of mass and $t_{0}$ was set as the time after which the cluster dynamics becomes stationary. These MSD curves are well fitted by the active Brownian particle model:
${\left< \Delta r^{2}\right> = 2V_{c}^{2}\tau_{p}\left [ t - \tau_{p}\left ( 1 - e^{-t/\tau_{p}} \right ) \right ]}$ 
\cite{fily2012athermal,fily2014freezing,marchetti2016minimal,villa2020run},
where $V_c$ is the effective cluster self-propulsion speed, $\tau_p$ is its persistence time and we relabelled $t-t_0$ as just $t$. Ballistic motion occurs at short times ($t \ll \tau_p$), where $\left< \Delta r^{2}\right> \approx V_c^2 t^2$, crossing over to normal diffusion at long times ($t \gg \tau_p$), where $\left< \Delta r^{2}\right> \approx 2 V_c^2 \tau_p t$.

For $J=0$ (Fig.~\ref{fig2}b), $\langle \Delta r^{2}\rangle$ decreases with cluster mass in both ballistic and diffusive regimes. Without alignment, particle polarities often oppose each other, leading to a decrease of cluster speed with mass.
 The persistence time $\tau_p$ equals the single-particle value $\tau_R$ and is mass-independent since for $J=0$ reorientations are unaffected by particle-particle interactions. 
Near the flocking transition ($J \approx 0.6$, Fig.~\ref{fig2}c), $V_c$ still decreases with mass, while alignment suppresses fluctuations and makes the cluster persistence time $\tau_p$ increase with mass. These opposing trends compensate each other, yielding convergence of all $\langle \Delta r^{2}\rangle(t)$ curves at long times, where they become mass-independent.
Above the transition ($J=10$, Fig.~\ref{fig2}d), particles align their polarities, and clusters move coherently at nearly $V_c=v_0$, making $V_c$ mass-independent. Larger clusters retain longer $\tau_p$, so the long-time MSD now increases with mass.

\begin{figure}[!h]
\centering
\includegraphics[width=1.0\columnwidth]{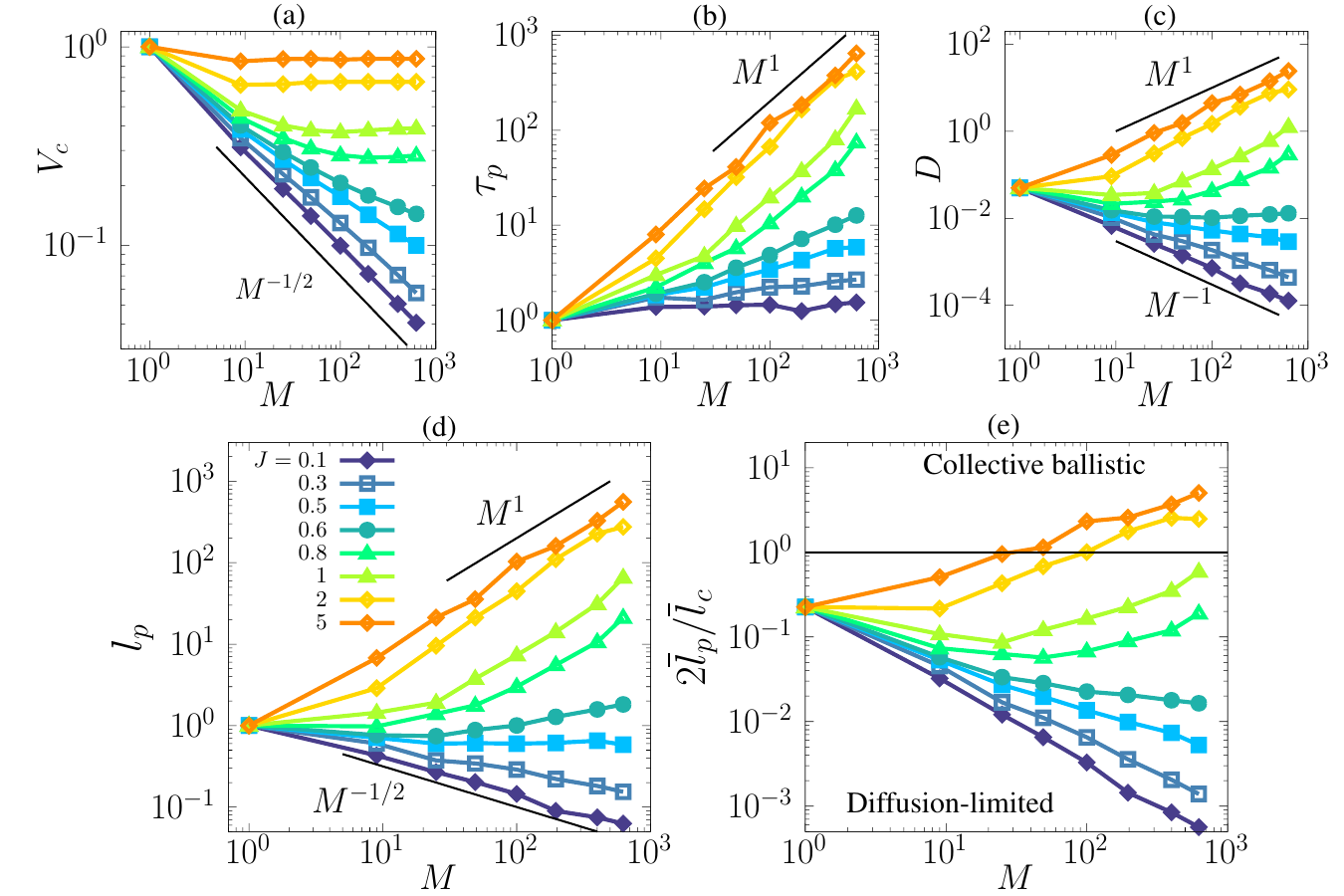}
\caption{Self-alignment fundamentally alters how cluster properties depend on mass, explaining aggregation kinetics regimes. For different  \( J \), the dependence on cluster mass, obtained from single-cluster simulations, is shown for
(a) cluster speed \( V_c \), (b) cluster persistence time \( \tau_p \), (c) diffusion coefficient \( D = V_c^2 \tau_p / 2 \), showing transition from mass-hindered to mass-enhanced diffusion,  
(d) persistence length \({ l_p = V_c \tau_p }\), and  
(e) kinetic criterion ratio \( 2\overline{l}_p / \overline{l}_c \), where we estimate $\overline{l}_{c}$ from the global density. A crossover to ballistic aggregation occurs  when \( 2\overline{l}_p / \overline{l}_c > 1 \).
}

\label{fig3}
\end{figure}

We quantify the mass-dependence of the effective active cluster motion parameters obtained from single-cluster simulations. Figure~\ref{fig3}a shows $V_c$ versus mass for varying $J$. For small $J$,  $V_c \sim M^{\gamma}$ with $\gamma \simeq -1/2$, as expected for the magnitude of the average of $M$ random vectors. As $J$ increases, $\gamma$ rises to $\gamma = 0$, giving $V_c=v_0$.
Figure~\ref{fig3}b shows the cluster persistence time $\tau_p$ versus mass. We see that $\tau_p$ grows nearly as $\tau_p \sim M^{\nu}$, with $\nu$ increasing from $0$ to approximately $1$. The long-time diffusion coefficient is $D  = V_c^2 \tau_p/2$, so  $D \sim M^{\alpha}$, where $\alpha = 2\gamma + \nu$. 
Fig.~\ref{fig3}c shows that $-1 \leq \alpha \leq 1$. For low (high) $J$, $D$ decreases (increases) with mass. Notice that, for extreme values of $J$, we observe well-defined power-law-like dependencies on $M$. At intermediate $J$, these cluster motion parameters retain a monotonic dependence on $M$. Thus, we will assume power-law dependencies in our theory below in order to capture the limiting values of $z$, as well as its trend, as $J$ is varied.

To characterize the aggregation regime, we compare the average intercluster distance, $\overline{l}_{c}$, with the cluster persistence length, $l_{p} = V_{c}\,\tau_{p}$; also, see Fig.~\ref{fig1}b. If clusters collide and merge before traveling a persistence length, the aggregation is estimated to occur in the ballistic regime. Conversely, if clusters travel a distance larger than $l_{p}$ before merging, the process is DLCA. 
The average intercluster distance $\overline{l}_{c}$ is estimated from the global particle density $\rho = N/L^2$ and the average cluster mass $\overline{M}$ via $\overline{l}_{c} = \sqrt{\overline{M}/\rho}$ \cite{beatrici2017mean}.

Figure~\ref{fig3}d shows that $l_{p}(M) \sim M^{\delta }$, with $\delta = \nu + \gamma$. Applying for $M=\overline{M}$, we define $\overline{l}_{p}(\overline{M}) = l_{p}(\overline{M})\sim \overline{M}^{\delta }$
which we then compare to $\overline{l}_{c}$
to identify the aggregation regime. 
If the clusters are initially separated by a distance \( l_c \), the simplified condition for ballistic collision reads \( 2l_p / l_c > 1 \). 
Fig.~\ref{fig3}e shows $2\overline{l}_p / \overline{l}_c$ versus $\overline{M}$.
For \( J \geq 2 \), the curves cross the criterion $2\overline{l}_p / \overline{l}_c=1$ at \( \overline{M} \approx 10^2 \), indicating a ballistic aggregation regime for large masses consistent with the faster kinetics observed for \( \overline{M} \) in Fig.~\ref{fig1}a. For lower, intermediate values of $J$ in the range \( 0.8 \leq J \leq  1 \), the ratio $2\overline{l}_p / \overline{l}_c$ is expected to cross $1$ at larger \( \overline{M}\), suggesting that ballistic behavior emerges only at later times. This is also consistent with our measurements in Fig.~\ref{fig1}a, where the aggregation exponent of \( \overline{M} \) exhibits a crossover to a faster aggregation regime with \( z \simeq 2 \). 
In contrast, for \( J \leq 0.6 \), the ratio \( 2\overline{l}_p / \overline{l}_c \) remains below $1$ for all mass values, indicating that the system is in the DLCA regime throughout the whole aggregation process.

Therefore, the increase in \( z \) seen in Fig.~\ref{fig1}a signals a shift from DLCA to ballistic aggregation. This shift is driven by collective motion within clusters, which makes the persistence length grow linearly with the average mass while the ratio \( \overline{l}_p / \overline{l}_c \) scales as $\sim \overline{M}^{1/2}$. At sufficiently large $\overline{M}$, the ballistic aggregation criterion is satisfied. Notably, collective ballistic aggregation represents the asymptotic regime---not merely a transient effect.


As shall be important in our theory, we analyze the impact of $J$ on the full cluster mass distribution $P(M,t)$~\cite{de2021active}; see Supplemental Material \cite{supp}. 
These distributions show dynamical scaling: all curves collapse onto $f(M/\overline{M}) = \overline{M}^{2} P(M,t)$, consistent with irreversible aggregation scenarios~\cite{leyvraz2003scaling,kolb1984unified}.

\begin{figure}[!h]
\centering
\includegraphics[width=1.0\columnwidth]{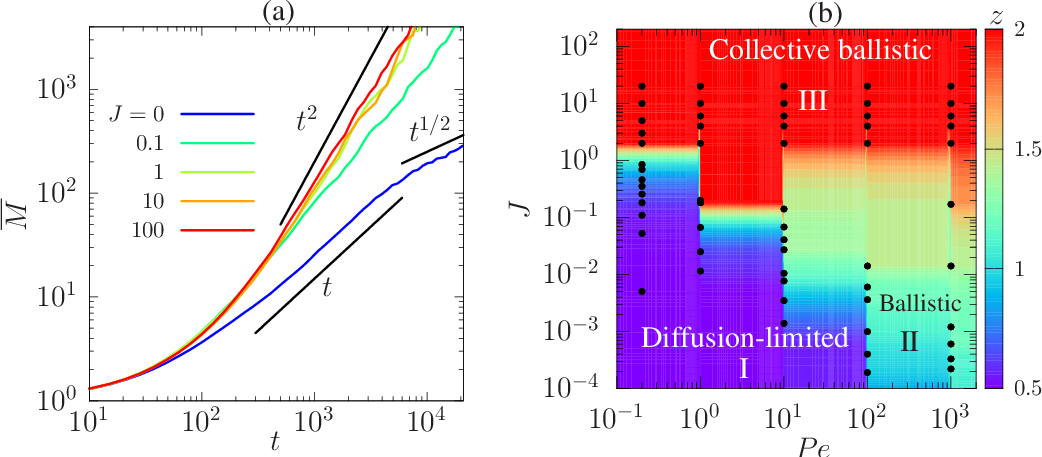}
\caption{High persistence enables non-collective ballistic aggregation with $z=1$, completing the aggregation regimes. (a) $\overline{M}(t) \sim t^z$ for different $J$ at $\mathrm{Pe} = 10^3$. For $J = 0$, one has $z \simeq 1$ (non-collective ballistic), a long-lived transient regime relevant in biological systems, eventually crossing over to DLCA. (b) Aggregation regime diagram in terms of the aggregation exponent $z$.
Region I (blue) is  DLCA; Region II (light blue) is ballistic aggregation; Region III (red) is collective ballistic aggregation (CBA).
 }
\label{fig4}
\end{figure}

To explain $z$ in each regime, we generalize Smoluchowski coagulation theory; see Supplementary Material~\cite{supp}. In the DLCA regime, we use an aggregation rate  \( K(M, M') \sim D(M) + D(M') \)~\cite{leyvraz2003scaling, moncho2004colloidal, moncho2000simulations, moncho2001dlca}. Considering the dynamical scaling of \(P(M,t)\), we obtain \( z = \frac{1}{1 - \alpha} \). Since \( -1 \leq \alpha \leq 0 \) for DLCA, this yields \( \frac{1}{2} \leq z \leq 1 \), matching our simulations.
A similar approach applies to ballistic aggregation. We employ the aggregation rate kernel,
\(
K(M, M') \sim \, \left| V_{c}(M) - V_{c}(M') \right| \, (R(M) + R(M')),
\) 
where \( R(M) \sim M^{1/2} \) is the average cluster radius~\cite{leyvraz2003scaling}. We obtain
\(
{z = \frac{2}{1 - 2\gamma}}
\) \cite{supp}. For \( J \geq 0.8 \),  ballistic aggregation occurs with mass-independent cluster speed (\( \gamma = 0 \)), yelding \( z = 2 \), consistent with our simulations (Fig.~\ref{fig4}a).

{\it High particle persistence and diagram}---For high particle persistence, $\mathrm{Pe}=10^3$, and $J = 0$, non-collective ballistic motion emerges. Fig.~\ref{fig4}a shows $z \simeq 1$. As in the previous case of low $\mathrm{Pe}$ and $J = 0$ (Fig.~\ref{fig3}a-b), we observed $V_c \sim M^{-1/2}$ and mass-independent $\tau_p=\tau_R$. A high $\mathrm{Pe}$ does not affect particle alignments, giving $\overline{l}_p =V_c \tau_p \sim 10^3/\sqrt{\overline{M}}$ and $2\overline{l}_p/\overline{l}_c \simeq 225.73\overline{M}^{\,-1}$ at our density. Thus, $\overline{M} < 225.73$ yields ballistic aggregation, while larger masses follow DLCA. This initial power-law during ballistic aggregation is also important as long-lived transients are relevant in biological systems due to the not-so-large number of particles. Smoluchowski theory~\cite{supp} gives $z=1$, matching simulations.

At high $\mathrm{Pe}$, increasing $J$ yields again ${l_p \sim M}$ and consequently ${2\overline{l}_p/\overline{l}_c \sim \overline{M}^{1/2} > 1}$, indicating ballistic aggregation. With mass-independent cluster speed ($\gamma = 0$, Fig.~\ref{fig3}a), Smoluchowski theory predicts $z = 2$, matching our simulations once again. Therefore, \textit{collective ballistic aggregation} (CBA) occurs for both low and high persistence.

Figure~\ref{fig4}b summarizes the aggregation regimes in the $\mathrm{Pe}$--$J$ plane. 
At low $J$ and $\mathrm{Pe}$ the system follows diffusion-limited cluster aggregation (Region~I); 
increasing $\mathrm{Pe}$ at low $J$ leads to ballistic aggregation driven by persistence without 
collective effects (Region~II); and at high $J$, flocking within clusters produces coherent motion 
characteristic of collective ballistic aggregation (Region~III).

{\it Conclusions}---We presented a comprehensive study elucidating how collective motion enables fast cluster aggregation in adhesive active matter. Varying alignment strength (\( J \)) and Péclet number (\( \mathrm{Pe} \)), we identified distinct regimes with scaling  \( \overline{M}(t) \sim t^z \), where \( z \in [1/2,2] \). Single-cluster analysis revealed how the persistence length \( l_p \sim M^{\delta} \) emerges from the interplay between alignment and cluster size, explaining $z$.

At low \(J\), clusters diffuse with an effective diffusivity that decreases with cluster mass, consistent with DLCA~\cite{nakajima2011kinetics,durand2021}. Simulations and Smoluchowski theory yield \(z \in [1/2,1]\), matching classical DLCA. Increasing \(J\)  enhances internal cluster alignment, triggering a ballistic aggregation that eventually gives \(2l_p/l_c>1\), yielding \(z\approx 2\) in the high-$J$ limit where cluster speed becomes independent of mass ($\gamma=0$). 
Therefore, we identify a mechanism that allows for a \textit{collective ballistic aggregation} regime which is asymptotic (not transient) and driven by internal alignment rather than individual persistence. Our results offer a natural explanation for the anomalous aggregation reported in prior studies~\cite{mones2015anomalous,cochet2017physical,mehes2012collective}. Moreover,
increasing \( \mathrm{Pe} \) reveals that even without alignment (\( J = 0 \)), high persistence leads to a long-lived ballistic aggregation transient, relevant to biological systems, which eventually undergoes a crossover to DLCA. After validating our simulation results through theory, we mapped all regimes in a $\mathrm{Pe}$--$J$ diagram (Fig. \ref{fig4}b).

Our framework elucidates how alignment and persistence govern aggregation dynamics in adhesive active matter and offers predictive insights for biological processes like tissue formation, where mass-dependent collective behavior is crucial. It also explains the diversity of experimental aggregation exponents. Future work should aim to use more detailed experiments to determine the model parameters and further validate our results; to investigate high-density and cluster-fragmentation effects; and to explore how medium properties (e.g., viscoelasticity~\cite{beysens2000cell}) or contact inhibition~\cite{bothe2025eph} modulate kinetics.

\begin{acknowledgments}
  We thank the Brazilian agencies CAPES, CNPq, FAPERGS and FAPESP. H.C.M.F.\ and L.G.B.\ acknowledge CNPq (procs.\ 402487/2023-0 and 443517/2023-1). E.F.T.\ acknowledges ICTP-SAIFR/IFT-UNESP. The simulations used the \href{https://pnipe.mcti.gov.br/laboratory/19775}{VD Lab} cluster at IF-UFRGS. P.d.C.\ was supported by Scholarships No.\ 2021/10139-2 and No.\ 2022/13872-5 and ICTP-SAIFR Grant No.\ 2021/14335-0, all granted by São Paulo Research Foundation (FAPESP), Brazil.
\end{acknowledgments}
\bibliographystyle{apsrev4-2}
\bibliography{references}


\clearpage
\onecolumngrid

\begin{center}
  \textbf{\large Supplementary Material: Collective ballistic motion explains fast aggregation in adhesive active matter}
\end{center}

\setcounter{equation}{0}
\setcounter{figure}{0}
\setcounter{table}{0}
\setcounter{page}{1}
\makeatletter
\renewcommand{\theequation}{S\arabic{equation}}
\renewcommand{\thefigure}{S\arabic{figure}}
\renewcommand{\bibnumfmt}[1]{[S#1]}
\renewcommand{\citenumfont}[1]{S#1}

\section{Additional Simulation Details}
We use $k_c = 200v_0/(\sigma\mu)$, $k_{\rm adh} = 10v_0/(\sigma\mu)$, $v_{0} = 0.1$,  $\mu = 1$, $\sigma = 1$ and $l_{\rm adh} = 1.5\sigma$.  Thus, particles irreversibly bind when they meet: the lowest adhesion, $k_{\rm adh}\sigma$, is $10$ times the self-propulsion, $v_0/\mu$, so particles move only in directions that do not destroy binding. We employ the Euler-Maruyama algorithm with a time step $\Delta t = 0.001\tau_0$.

\section*{Cluster Mass Distributions and Dynamical Scaling}
We present here our results for the effect of the alignment strength \( J \) on the cluster mass distribution \( P(M,t) \) in the low (\( \mathrm{Pe} = 1 \)) and high Péclet number regimes (\( \mathrm{Pe} = 10^3 \)). 
\begingroup
  \renewcommand{\thefigure}{S\arabic{figure}} 
\begin{figure}[!h]
\centering
\includegraphics[width=0.6\columnwidth]{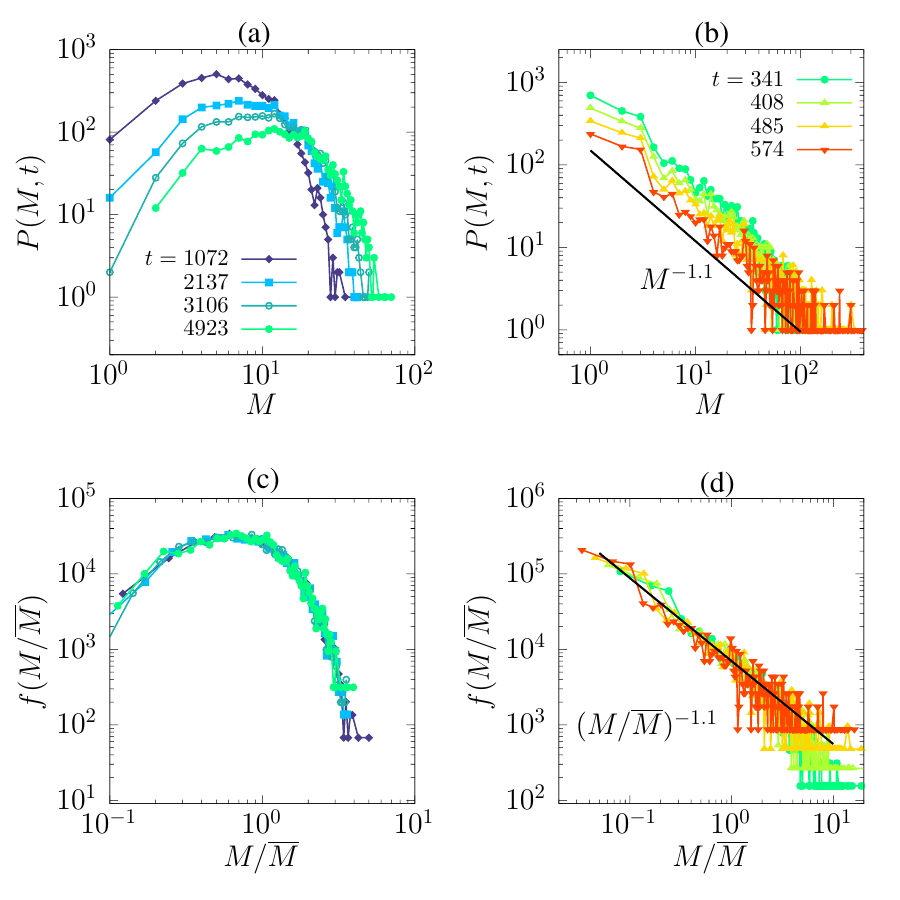}
\caption{
Cluster mass distribution \( P(M,t) \) in the high Péclet number regime (\( \mathrm{Pe} = 10^3 \)).  
(a) For \( J = 0 \), below the flocking threshold, the distribution has a peak, reflecting a characteristic cluster size.  
(b) For \( J = 100 \), above the flocking transition, the distribution becomes power-law, \( P(M,t) \sim M^{-\lambda} \), with \( \lambda \simeq 1.1 \). For (c) $J=0$ and (d) $J=100$, the rescaled distributions collapse onto a single curve, \( f(M/\overline{M}) = \overline{M}^2 P(M,t) \), confirming dynamical scaling \cite{leyvraz2003scaling}.
}
\label{fig:figS1}
\end{figure}
\endgroup

\begingroup
  \renewcommand{\thefigure}{S\arabic{figure}} 
\begin{figure}[!h]
\centering
\includegraphics[width=0.6\columnwidth]{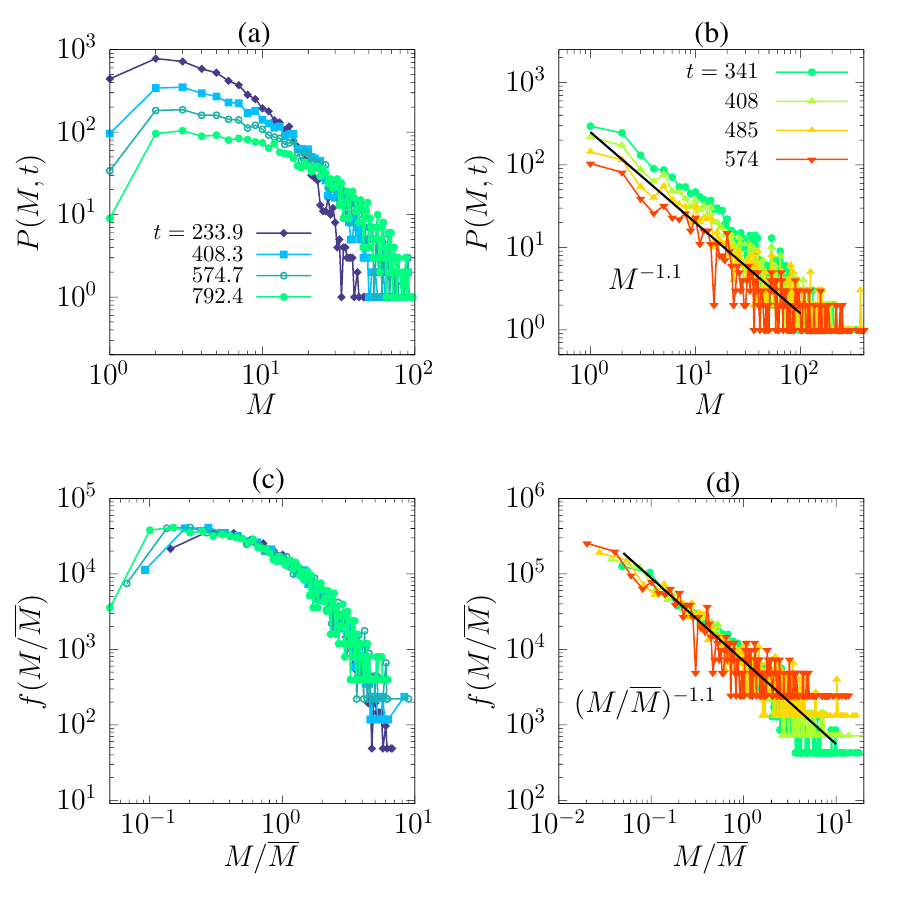}
\caption{
Similar to Figure~\ref{fig:figS1}, now in the high Péclet number regime (\( \mathrm{Pe} = 10^3 \)).
}
\label{fig:figS2}
\end{figure}
\endgroup
Figures~\ref{fig:figS1}a and b show \( P(M,t) \) for different time instants after the system has entered the algebraic coarsening regime \cite{leyvraz2003scaling}, for low Pe. For \( J = 0 \), below the flocking transition, the distribution exhibits a non-monotonic shape, indicating the presence of a characteristic cluster size. In contrast, for strong alignment (\( J = 100 \)), the distribution develops a power-law tail, \( P(M,t) \sim M^{-\lambda} \), with exponent \( \lambda \simeq 1.1 \), consistent with scale-free cluster growth driven by collective motion. We observe in Figures~\ref{fig:figS1}c and d that the distributions obey dynamical scaling. When rescaled by the mean cluster mass \( \overline{M}(t) \), all curves collapse onto a single universal function,
\(
f\left( \frac{M}{\overline{M}} \right) = \overline{M}^2 P(M,t),
\)
demonstrating self-similar evolution during the coarsening process. The same universal scaling function was observed in Ref.~\cite{leyvraz2003scaling} for another irreversible aggregation problem. For high Pe, the behavior of the cluster mass distributions is similar; see Figure~\ref{fig:figS2}.

\section{Cluster aggregation theory: The Smoluchowski coagulation equation}
We present here a cluster aggregation theory based on the Smoluchowski coagulation equation \cite{smoluchowski1916drei} generalized to the types of mass-dependent cluster movements of our problem. This theoretical approach builds on the derivation presented in Ref.~\cite{leyvraz2003scaling}, incorporating modifications where appropriate. For completeness, we include intermediate steps common to other irreversible aggregation problems. Consider clusters with different masses moving through space. When two clusters of masses \( M \) and \( M' \) come sufficiently close, they may irreversibly merge into a single cluster of mass \( M + M' \):
\begin{eqnarray}
     A_{M} + A_{M'} \xrightarrow[K\left ( M,M' \right )]{} A_{M+M'}.
\end{eqnarray}
Here, $A_M$ denotes a cluster and \( K(M, M') \) denotes the rate at which two clusters of masses \( M \) and \( M' \) coagulate. This quantity is known as the coagulation kernel.

We assume a continuous mean-field approach characterized by spatial homogeneity, where the cluster mass distribution is uniform throughout the entire space:
\begin{eqnarray}
     \int_{V_{0}} P(\mathbf{r},M,t) \, dV = P(M,t) V_{0},
\end{eqnarray}
where $P(M,t)$ is the spatially-homogeneous cluster mass distribution and $V_{0} = L^{d}$ is the total $d$-dimensional hypercubic volume. Notice that, while our simulations were performed in $d=2$, consistent with many experimental setups, we formulate the theory in arbitrary dimension $d$, as the generalization is straightforward.
We now define the concentration of the mass distribution as
\begin{eqnarray}
    C(M,t) = \frac{P(M,t)}{V_{0}}.
\end{eqnarray}

The concentration \( C(M,t) \) increases when two smaller clusters of masses \( M' \) and \( M - M' \) merge to form a cluster of mass \( M \). The rate of this process is proportional to the product \( C(M',t)\, C(M - M',t) \). On the other hand, when a cluster of mass \( M \) merges with any other cluster, it is removed from the population, resulting in a decrease in \( C(M,t) \). To describe the full aggregation process, we integrate over the coagulating masses, treated as continuous variables:
\begin{eqnarray}
    \label{1}
    \frac{\partial C(M,t)}{\partial t} = \frac{1}{2} \int_0^M K(M', M - M')\, C(M',t)\, C(M - M',t) \, dM'  - \int_0^\infty K(M, M')\, C(M,t)\, C(M',t) \, dM',
\end{eqnarray}
where \( K(M, M') \) is the coagulation kernel. The prefactor \( 1/2 \) ensures that the contribution is not double-counted. Equation~\eqref{1} is the Smoluchowski coagulation equation~\cite{smoluchowski1916drei}. In terms of the mass distribution \( P(M,t) \), Eq.~\eqref{1} becomes
\begin{eqnarray}
    \label{eq_smo}
    \frac{\partial P(M,t)}{\partial t} = \frac{1}{2\,V_{0}} \int_0^M K(M', M - M')\, P(M',t)\, P(M - M',t) \, dM' - \frac{1}{V_{0}}\int_0^\infty K(M, M')\, P(M,t)\, P(M',t) \, dM'.
\end{eqnarray}

We note that the total mass is conserved:
\begin{eqnarray}
    \label{2}
    M_0 = \int_0^\infty M\, P(M,t) \, dM = \text{const}.
\end{eqnarray}
Also, we define the total number of clusters:
\begin{eqnarray}
    \label{3}
    N(t) = \int_0^\infty P(M,t) \, dM.
\end{eqnarray}
Using Eq.~\eqref{eq_smo}, we compute the time derivative of \( N(t) \):
\begin{eqnarray}
   \label{5}
   \frac{dN(t)}{dt} &=& \frac{1}{2\,V_{0}} \int_0^\infty  \int_0^M K(M', M - M') \, P(M',t)\, P(M - M',t) \, dM' \, dM \nonumber \\
   && - \frac{1}{V_{0}}\int_0^\infty \int_0^\infty K(M, M') \, P(M,t)\, P(M',t) \, dM' \, dM.
\end{eqnarray}
We now interchange the order of integration in the first term and perform the change of variables \( M'' = M - M' \):
\begin{eqnarray}
   \frac{dN(t)}{dt} &=& \frac{1}{2\,V_{0}} \int_0^\infty \int_0^\infty K(M', M'') P(M',t) P(M'',t) \, dM'' \, dM' \nonumber \\
   && - \frac{1}{V_{0}} \int_0^\infty \int_0^\infty K(M, M') P(M,t) P(M',t) \, dM' \, dM.
\end{eqnarray}
Renaming dummy variables \( M' \to M \) and \( M'' \to M' \), this becomes
\begin{eqnarray}
   \frac{dN(t)}{dt} = -\frac{1}{2\,V_{0}} \int_0^\infty \int_0^\infty K(M, M') P(M,t) P(M',t) \, dM' \, dM.
   \label{6}
\end{eqnarray}
The average cluster mass is defined as
\begin{eqnarray}
    \label{7}
     \overline{M}(t) = \frac{\int_0^\infty M P(M,t) \, dM}{\int_0^\infty P(M,t) \, dM} = \frac{M_0}{N(t)}.
\end{eqnarray}
Thus,
\begin{eqnarray}
    \label{8}
    N(t) = \frac{M_0}{\overline{M}(t)}.
\end{eqnarray}
Taking the time derivative yields
\begin{eqnarray}
     \label{9}
    \frac{dN(t)}{dt} = -\frac{M_0}{\overline{M}(t)^2} \frac{d\overline{M}(t)}{dt}.
\end{eqnarray}
Substituting Eq.~\eqref{9} into Eq.~\eqref{6} leads to
\begin{eqnarray}
     \label{10}
    \frac{d\overline{M}(t)}{dt} = \frac{\overline{M}(t)^2}{2\,M_0\,V_0} \int_0^\infty \int_0^\infty K(M, M') P(M,t) P(M',t) \, dM' \, dM.
\end{eqnarray}

\subsection{Diffusion-limited cluster aggregation}

The physical processes governing the rate of cluster coagulation vary depending on the specific scenario. We first focus on diffusion-limited aggregation, where cluster diffusion governs the coagulation rate. Other factors such as cluster shape and collision cross-section can also be incorporated. For a sphere of radius \( R \) surrounded by Brownian particles with diffusion constant \( D \), the absorption rate scales as \( K \sim D R^{d-2} \)~\cite{krapivsky2010kinetic}. Based on this idea, a widely used kernel for DLCA~\cite{leyvraz2003scaling,moncho2004colloidal,moncho2000simulations,moncho2001dlca} is
\begin{eqnarray}
   \label{11}
K(M,M') = A \left[ D(M) + D(M') \right] \left[ R(M) + R(M') \right]^{d-2} = A \left( M^\alpha + M'^\alpha \right) \left( M^{1/d_f} + M'^{1/d_f} \right)^{d-2},
\end{eqnarray}
where \( A \) is a constant that ensures correct dimensionality (volume/time), \( R \sim M^{1/d_f} \) is the radius of gyration with fractal dimension \( d_f \), and \( D(M) \sim M^\alpha \) is the diffusion-mass scaling (observed in our simulations here and in other references such as Ref.~\cite{beatrici2017mean}). Unlike fixed spheres absorbing particles, here clusters themselves diffuse and merge, thus the kernel includes sums of diffusion constants and radii.

Substituting Eq.~\eqref{11} into Eq.~\eqref{10}, we get
\begin{eqnarray}
   \label{12}
    \frac{d\overline{M}(t)}{dt} = \frac{A\, \overline{M}(t)^2}{2\,M_0\,V_0} \int_0^\infty \int_0^\infty \left( M^\alpha + M'^\alpha \right) \left( M^{1/d_f} + M'^{1/d_f} \right)^{d-2} P(M,t) P(M',t) \, dM' \, dM.
\end{eqnarray}
From our simulations, as discussed above, we observe the dynamical scaling form \cite{leyvraz2003scaling}
\begin{eqnarray}
    \label{17}
    P(M,t) = \overline{M}(t)^{-2} f\left( \frac{M}{\overline{M}(t)} \right).
\end{eqnarray}
Applying the change of variables \( x = M / \overline{M}(t) \) and \( x' = M' / \overline{M}(t) \), Eq.~\eqref{12} becomes
\begin{eqnarray}
    \frac{d\overline{M}(t)}{dt} &=& \frac{A \overline{M}(t)^{\alpha + \frac{d-2}{d_f}}}{2 M_0 V_0} \int_0^\infty \int_0^\infty (x^\alpha + x'^\alpha)(x^{1/d_f} + x'^{1/d_f})^{d-2} f(x) f(x') \, dx \, dx' \nonumber \\
    \label{22}
    &=& C_1 \overline{M}(t)^{\alpha + \frac{d-2}{d_f}},
\end{eqnarray}
where
\begin{eqnarray}
    I_1 = \int_0^\infty \int_0^\infty (x^\alpha + x'^\alpha)(x^{1/d_f} + x'^{1/d_f})^{d-2} f(x) f(x') \, dx \, dx'
\end{eqnarray}
is a constant, and
\begin{eqnarray}
    C_1 = \frac{A I_1}{2 M_0 V_0}.
\end{eqnarray}
Integrating Eq.~\eqref{22} with initial condition \(\overline{M}(0) = 1\) yields
\begin{eqnarray}
     \overline{M}(t) = \left[ C_1 \left( 1 - \alpha - \frac{d-2}{d_f} \right) t + 1 \right]^{\frac{d_f}{d_f - d_f \alpha - (d-2)}} \sim t^z,
\end{eqnarray}
with the dynamic exponent
\begin{eqnarray}
    \label{23}
    z = \frac{d_f}{d_f - d_f \alpha - (d-2)}.
\end{eqnarray}
In two dimensions (\( d=2 \)), this simplifies to
\begin{eqnarray}
    z = \frac{1}{1 - \alpha}.
\end{eqnarray}

Hence, the average cluster mass grows algebraically in time as a direct consequence of the dynamical scaling hypothesis.

\subsection{Ballistic aggregation}

In ballistic aggregation, cluster collisions are dominated by their relative velocity. A commonly used kernel in the literature is~\cite{jiang1993scaling,leyvraz2003scaling}
\begin{eqnarray}
     \label{24}
     K(M,M') = B \left| V(M) - V(M') \right| \left( R(M) + R(M') \right)^{d-1} = B \left| M^\gamma - M'^\gamma \right| \left( M^{1/d_f} + M'^{1/d_f} \right)^{d-1}.
\end{eqnarray}
Here, \( |V(M) - V(M')| \) originates from the relative velocity between clusters and the term \( (R(M) + R(M'))^{d-1} \) corresponds to their collision cross-section. As we observed in our simulations, the typical cluster speed scales with mass as \( V \sim M^\gamma \). For clusters composed of particles with uncorrelated self-propulsions, \( \gamma = -1/2 \), whereas for fully aligned clusters, \( \gamma = 0 \).

Substituting Eq.~\eqref{24} into Eq.~\eqref{10} yields
\begin{eqnarray}
     \label{25}
     \frac{d\overline{M}(t)}{dt} = \frac{B \overline{M}(t)^2}{2 M_0 V_0} \int_0^\infty \int_0^\infty \left| M^\gamma - M'^\gamma \right| \left( M^{1/d_f} + M'^{1/d_f} \right)^{d-1} P(M,t) P(M',t) \, dM' \, dM.
\end{eqnarray}
Applying the scaling ansatz Eq.~\eqref{17}, and changing variables as before, leads to
\begin{eqnarray}
    \frac{d\overline{M}(t)}{dt} &=& \frac{B \overline{M}(t)^{\gamma + \frac{d-1}{d_f}}}{2 M_0 V_0} \int_0^\infty \int_0^\infty |x^\gamma - x'^\gamma| (x^{1/d_f} + x'^{1/d_f})^{d-1} f(x) f(x') \, dx \, dx' \nonumber \\
    \label{27}
    &=& C_2 \overline{M}(t)^{\gamma + \frac{d-1}{d_f}},
\end{eqnarray}
where
\begin{eqnarray}
    I_2 = \int_0^\infty \int_0^\infty |x^\gamma - x'^\gamma| (x^{1/d_f} + x'^{1/d_f})^{d-1} f(x) f(x') \, dx \, dx'
\end{eqnarray}
is a constant and
\begin{eqnarray}
    C_2 = \frac{B I_2}{2 M_0 V_0}.
\end{eqnarray}
Integrating Eq.~\eqref{27} with \(\overline{M}(0) = 1\) gives
\begin{eqnarray}
    \overline{M}(t) = \left[ C_2 \left( 1 - \gamma - \frac{d-1}{d_f} \right) t + 1 \right]^{\frac{d_f}{d_f - d_f \gamma - (d-1)}} \sim t^z,
\end{eqnarray}
where the dynamic exponent is
\begin{eqnarray}
    \label{28}
    z = \frac{d_f}{d_f - d_f \gamma - (d-1)}.
\end{eqnarray}
Assuming compact, non-fractal clusters where the spatial dimension equals the fractal dimension, \( d = d_f \), the dynamic exponent reduces to
\begin{eqnarray}
    \label{29}
    z = \frac{d}{1 - d\gamma}.
\end{eqnarray}

Again, the average cluster mass grows algebraically in time with the exponent \( z \) depending on the scaling of speed, fractal dimension, and system dimension. For $d=2$, we obtain $z = \frac{2}{1 - 2\gamma}$ as discussed in the main text. In particular, for non-collective ballistic aggregation, one has $\gamma=-1/2$ and we obtain $z=1$.

\section*{Movies}
For illustration, \textbf{Movie 1} shows the aggregation for (a) $J=100$ and $\mathrm{Pe}=1$, (b) $J=100$ and $\mathrm{Pe}=1000$, (c) $J=0$ and $\mathrm{Pe}=1$, and (d) $J=0$ and $\mathrm{Pe}=1000$. To facilitate visualization, we used $\phi=0.1$. Colors indicate the polarization of the particles; for a color wheel, see main text. \textbf{Movie 2} shows single-cluster simulations for the same values of $J$ and $\mathrm{Pe}$.

The films reveal that cluster fractality becomes more pronounced at higher densities and extraordinary values of $J$. To observe fractality in the exponents $z$ measured in our main text figures, simulations at exceptional scales of mass and time would be required. To verify that these effects indeed emerge at large masses, we modified the averaging procedure for the average cluster mass in order to weight larger clusters more strongly, and observed a slight increase in $z$ that is consistent with the theory above.

\end{document}